\documentclass[12pt]{article}
\usepackage{amsfonts}
\usepackage{amssymb}
\usepackage{amsmath}
\usepackage{graphicx}
\usepackage{psfrag}
\usepackage{epsfig}

\renewcommand{\baselinestretch}{1.1}

\def\IR{\mathbb{R}}
\def\IC{\mathbb{C}}
\def\IZ{\mathbb{Z}}

\def\Tr{\mbox{Tr}}

\def\id{\mbox{id}}

\def\be{\begin{equation}}
\def\ee{\end{equation}}
\def\bea{\begin{eqnarray}}
\def\eea{\end{eqnarray}}

\def\Div{{\rm Div}}

\def\ibar{{\bar\imath}}
\def\jbar{{\bar\jmath}}

\def\id{\protect{{1 \kern-.28em {\rm l}}}}

\makeatletter
\renewcommand\section{\@startsection {section}{1}{\z@}%
                                   {-3.5ex \@plus -1ex \@minus -.2ex}%
                                   {2.3ex \@plus.2ex}%
                                   {\normalfont\large\bfseries}}

\renewcommand\subsection{\@startsection{subsection}{2}{\z@}%
                                   {-3.25ex\@plus -1ex \@minus -.2ex}%
                                   {1.5ex \@plus .2ex}%
       {\normalfont\normalsize\bfseries}}
\makeatother

\begin{document}

\begin{titlepage}

\begin{flushright}
hep-th/0509132\\
PUPT-2174\\
ITEP-TH-56/05\\
\end{flushright}

\vspace{5mm}

\begin{center}
{\huge Perturbative Gauge Theory}
\\
\vspace{3mm}
{\huge and
Closed String Tachyons
}\\
\vspace{1mm}
\end{center}

\vspace{1mm}
\begin{center}
{\large A. Dymarsky${}^1$, I.R. Klebanov${}^1$ and R. Roiban${}^{1,2}$
}\\
\vspace{3mm}
${}^1$
Joseph Henry Laboratories, Princeton University,
 Princeton, NJ  08544, USA
\vspace{2mm}

${}^2$
Physics Department, Pennsylvania State University, University Park,
PA 16802, USA

\vspace{2mm}
\end{center}

\vspace{3mm}

\begin{center}
{\large Abstract}
\end{center}
We find an interesting connection between perturbative large $N$
gauge theory and closed superstrings. The gauge theory in
question is found on $N$ D3-branes placed at the tip of
the cone $\IR^6/\Gamma$. In our previous work we showed that, when the 
orbifold group $\Gamma$ breaks all supersymmetry, then typically
the gauge theory is not conformal because of
double-trace couplings whose one-loop beta functions do not possess real zeros.
In this paper we observe a precise correspondence between
the instabilities caused by
the flow of these double-trace couplings and
the presence of
tachyons in the twisted sectors of type IIB theory on orbifolds
$\IR^{3,1}\times \IR^6/\Gamma$.
For each twisted sectors that does not contain tachyons,
we show that the corresponding double-trace coupling flows to a fixed point
and does not cause an instability. However, whenever a twisted sector
is tachyonic, we find that the corresponding one-loop beta function
does not have a real zero, hence an instability
is likely to exist in the gauge theory.
We demonstrate explicitly the one-to-one correspondence between
the regions of stability/instability in the space of charges under $\Gamma$
that arise
in the perturbative gauge theory and in the free string theory.
Possible implications of this remarkably simple gauge/string
correspondence are discussed.

\noindent
\vfil
\begin{flushleft}
2005
\end{flushleft}
\vfil
\end{titlepage}

\newpage

\renewcommand{\baselinestretch}{1.1}  

\section{Introduction}

A class of generalizations of the
maximally supersymmetric AdS$_{5}$/CFT$_4$ duality
\cite{jthroat,US,EW} (see \cite{MAGOO,Klebanov:2000me}
for reviews) is obtained through modding out both sides by a discrete
subgroup $\Gamma$ of the $SU(4)$ R-symmetry
\cite{KS,LNV}. In this fashion one obtains quiver gauge theories
with
${\cal N}=2$ supersymmetry for $\Gamma \subset SU(2)$,
with ${\cal N}=1$ supersymmetry for $\Gamma \subset SU(3)$,
and with no supersymmetry for all other $\Gamma$. In the large
$N$ limit,
correlation functions of the single-trace operators uncharged
under the orbifold group are ``inherited'' from the parent
${\cal N}=4$ SYM theory: they are the same up to a rescaling
of the `t Hooft coupling $\lambda =g_{\scriptscriptstyle YM}^2 N$
\cite{BKV,BJ}.
In particular, the beta functions for
marginal single-trace operators vanish in the large $N$
limit.
However, there is no such simple ``inheritance'' for
the twisted operators that are charged under the orbifold
group $\Gamma$.

The operators of particular interest to us
are double-trace operators $O^g O^{g^\dagger}$,
where $O^g$ is a twisted single-trace operator corresponding to the
element $g$ of the orbifold group $\Gamma$. Even though such operators
have no net charge, they are not protected by the
``inheritance principle'' \cite{BKV, BJ}.
Indeed, beta functions for marginal operators of this
type, made of scalar fields, are induced at one-loop order and are
found to be proportional to $\lambda^2$
\cite{TZ,Csaki,Adams}.
However, as shown in our previous work \cite{DKR}, this does
not necessarily imply that the large $N$ conformal invariance is lost:
an induced double-trace coupling may flow to a fixed point.

A systematic procedure for studying the issue of conformal invariance,
that was used in \cite{DKR}, is
to add each induced marginal double-trace operator to the action,
\begin{equation}
\delta S= - \int d^4 x f O\bar O  \ .
\end{equation}
The complete one-loop beta function in such a theory has the general
structure
\begin{equation} \label{genbeta}
M {\partial f\over \partial M} = \beta_f=  v_O f^2 + 2 \gamma_O \lambda f
+ a_O \lambda^2
\ .
\end{equation}
It is crucial that the right hand side  is not suppressed by powers of
$N$.
On the other hand, the beta function for $\lambda$ has no such
leading large $N$
contribution, due to the theorem of \cite{BKV,BJ}.\footnote{
Counting the powers of $N$, and using charge conservation,
 one can show that the double-trace
operators made of twisted single-trace ones
cannot induce any planar beta functions
for untwisted single-trace couplings.}
Therefore,
in the large $N$ limit, $\lambda$ may be dialed as we wish. In particular,
it can be made very small so that the one-loop approximation in
(\ref{genbeta}) is justified.
Then the equation $\beta_f=0$ has two solutions, $f= a_\pm \lambda$,
where
\begin{equation} \label{sol}
a_\pm ={1\over v_O} \left (-\gamma_O \pm \sqrt{ D }\right )\ ,
\qquad D= \gamma_O^2 - a_O v_O
\ .
\end{equation}

If the discriminant $D$ is non-negative, then
these solutions are real, so that $f$ may flow to an IR stable
fixed point.
If $D$ is negative, then there is no real fixed point
for $f$, and this coupling is not stabilized at least in the one-loop
approximation.
%
In the orbifold examples considered in \cite{DKR},
some double-trace operators were found to have $D\geq 0$, and hence
were
stabilized. Others were found to have a negative $D$, and hence caused
an instability of the weakly coupled gauge theory.
No non-supersymmetric orbifolds were found in \cite{DKR}
such that all double-trace couplings were stabilized, raising a
question whether any such gauge theories flow to a fixed line
passing through the origin.
It is natural to conjecture that the closed string
dual of this instability is associated with tachyonic modes from the
twisted sector of the orbifold $AdS_5\times S^5/\Gamma$ \cite{Adams,DKR}.
For a $\Gamma$ that does not act freely on the $S^5$, the tachyons are
present for all radii. Recalling that $R^4_{AdS}\sim  \alpha'^2\lambda$,
we then expect the instability in the dual field theory to be present for
all $\lambda$.
More interesting examples are
provided by the freely acting orbifolds where the twisted sector
closed strings are highly stretched at large $\lambda$, so that the
tachyons are lifted \cite{Adams}.
In \cite{DKR} it was suggested that, for $\lambda$ smaller than some
critical value, some closed string modes become tachyonic, causing an
instability that is manifested in the gauge theory
by the negative discriminant $D$ of the beta function.

Making such a correspondence between perturbative gauge
theory and closed string tachyons precise appears to be
a formidable task. The usual AdS/CFT correspondence seems to require
calculations in highly curved space with RR background fields, a problem
that has not yet been solved. In this paper we carry out a much simpler
calculation on the closed string side, computing the spectrum of twisted
closed strings on $\IR^{3,1}\times \IR^6/\Gamma$, i.e. in the limit
where the backreaction of the D3-branes is ignored. Remarkably, we find
that, whenever a given twisted sector contains a tachyon,
the discriminant $D$ of the beta function of the corresponding
double-trace coupling in the gauge theory is negative;
conversely, when the zero-point energy in a given twisted sector
vanishes, then $D$ is non-negative.\footnote{
A similar connection between closed string tachyons and instabilities of 
{\it non-commutative} gauge theories
on D3-brances at orbifold singularities was made in \cite{Armoni}.}
We demonstrate, for general orbifold groups $\Gamma$,
the one-to-one correspondence between
the regions of stability/instability in the space of charges under
$\Gamma$ that arise
in the perturbative gauge theory and in the free string theory.
This constitutes a remarkably simple correspondence between perturbative gauge
theory and free closed strings, which holds in the absence of
supersymmetry.
We speculate on reasons for why
this duality is at work in the Discussion section.
Using the results on the region of instability for double-trace couplings,
we prove that there are {\it no} non-supersymmetric
orbifold gauge theories possessing a fixed line passing through
the origin of the coupling constant space. This answers a
question posed in our previous paper \cite{DKR}.\footnote{ 
An investigation of other candidate non-supersymmetric large $N$
theories, such as those suggested in \cite{Frolov}, will be reported on
in a separate publication.}

On the way to deriving these main results, we also tie some loose ends
left over from our previous work \cite{DKR}.
We show that removing the $U(1)$ factors present in the $U(N)^k/U(1)$
quiver theories does not change the leading large $N$ beta function
calculations
carried out there using the orbifold projection methods. The only
modification is caused by the appearance of double-trace terms in the
tree-level action when the $U(1)$'s are removed.
In the case of the supersymmetry preserving orbifolds, we find that
for the operators induced at one-loop level the discriminant vanishes
%
%
and the double-trace beta functions have
the form $(f+ a\lambda)^2$, i.e. the coupling $f$ is stabilized at
$-a\lambda$. This coefficient of the double-trace term in the action is in
precise agreement with the value found through removing the $U(1)$'s
at tree level. This confirms the superconformal nature of the
supersymmetric $SU(N)^k$ orbifold gauge theories.

\section{Double-trace corrections for general orbifolds \label{gen_orb}}

In this section we review the results of \cite{DKR}
on the marginal double-trace operators in the 1-loop effective action of
orbifold gauge theories.
We will denote by $\Gamma\subset SU(4)$ the orbifold group
and by $g$ the representation of
the elements of $\Gamma$ in $SU(|\Gamma|N)$
where it acts by conjugation. The representation of $\Gamma$ in the
spinor and
vector representation of $SO(6)\sim SU(4)$ will be denoted by $r_g$ and
$R_{g}$, respectively. Since the vector representation of $SO(6)$ is the
2-index antisymmetric tensor representation of $SU(4)$, we have
$R_g=r_g\otimes r_g$.
The scalar fields invariant under the orbifold group satisfy
\begin{eqnarray}
\phi^I=R_g^{IJ}\,g\,\phi^J\,g^\dagger ~~.
\end{eqnarray}

\noindent
$\bullet$
The contribution of the fermion loop
to the double-trace part of the effective action is:
\begin{eqnarray}
\label{fermi_v1}
&&\delta S^{\rm 1~loop|2~tr}_{\rm Fermi}=\\
&&~~~~~~~~~ =\lambda^2\frac{\Div}{|\Gamma|} \sum_{g\in \Gamma} \,
\Tr[\gamma^I\gamma^J\gamma^K\gamma^Lr_g] \left[
2\Tr(\phi^J\phi^Ig^\dagger )\Tr(\phi^K\phi^Lg) +
\Tr(\phi^K\phi^Ig^\dagger )\Tr(\phi^J\phi^Lg)\right] \nonumber
\end{eqnarray}
where $\gamma^I$
denote the chiral (i.e. $4\times 4$) 6-dimensional Dirac
matrices.
Our notation for the divergent part of a generic
1-loop scalar amplitude is
\begin{eqnarray}
\Div=\int
\frac{d^{4}k}{(2\pi)^4}\,\frac{1}{k^4}=\frac{1}{16\pi^2}\ln
\frac{\Lambda^2}{M^2}~~,
\end{eqnarray}
where $\Lambda$ is the UV cutoff and $M$ is the renormalization
scale.

\noindent
$\bullet$ The contribution of the vectors, scalars and ghost loops
to the  double-trace part of the effective action is:
\begin{eqnarray}
\label{vs22}
\delta S^{\rm 1~loop|2~tr}_{\rm Bose,\,ghost}
\!\!&=&\!\!
-\lambda^2\frac{\Div}{2|\Gamma|}
\sum_{g\in \Gamma}
\Big\{
~~
\left(\Tr[R_g]+2\right)\left(\Tr(\phi^2 g^\dagger )\Tr(\phi^2 g)
+2\Tr(\phi^I\phi^J g^\dagger )\Tr(\phi^J\phi^I g) \right) \cr
&&+\vphantom{\Big|}
2(R_g^{KQ}+(R_g^{-1}){}^{KQ} )\Tr\left([\phi^I,\,\phi^Q]g^\dagger \right)
\Tr\left([\phi^I,\,\phi^K]\gamma\right)\\
&&-\vphantom{\Big|}
2\left(\delta^{KI}(R_g^{-1}){}^{PQ}+\delta^{PQ}R_g^{KI}
-2\delta^{PQ}\delta^{KI}
\right)\Tr\left(\phi^P\phi^Qg^\dagger \right)
\Tr\left(\phi^K\phi^Ig\right)~~\Big\}~~.
\nonumber
\end{eqnarray}

\noindent
$\bullet$
We see that double-trace operators
made out of twisted single-trace operators are
generated at 1-loop.
Perturbative renormalizability requires then that
they
be added to the tree level
action.
Such a tree-level deformation
\begin{eqnarray}
\delta_{\rm 2~trace}S={1\over
2}\sum_{g\in\Gamma}\,f_g\,O_g^{IJ}O_{g^\dagger }^{JI} ~~~~{\rm
with}~~~~O_g^{IJ}=\Tr(g\phi^I\phi^J)~~.
\label{def}
\end{eqnarray}
brings the following
additional contributions to the effective
action:
\begin{eqnarray}
\label{2trcontrib}
&&\delta S^{\rm 1~loop|2~tr}_{\rm 2~trace}
=-\frac{\Div}{|\Gamma|}\Big\{
\sum_{g\in \Gamma} f_g\left(\frac{1}{|\Gamma|}\sum_{\tilde g\in \Gamma}
f_{{\tilde g} g^\dagger  {\tilde g}^\dagger }\right)O_g^{IJ}O_{g^\dagger }^{JI}\\
\!\!&+&\!\!\lambda
\sum_{g\in\Gamma} f_g O^{JI}_{g^\dagger }
\left[
4\delta^{I{\hat I}}\delta^{J{\hat
J}}+\left(\delta^{IJ}+R_g^{JI}\right)\delta^{{\hat I}{\hat
J}}-2\left(\left(R_g^{-1}\right){}^{I{\hat I}}\delta^{J{\hat J}} +
\delta^{I{\hat I}} \left(R_g^{-1}\right){}^{J{\hat J}}\right)
\right]O_g^{{\hat I}{\hat J}}\Big\}~~.
\nonumber
\end{eqnarray}
One may recognize the bracket on the second line as the 1-loop
dilatation operator acting on twisted operators.

\noindent
$\bullet$
In \cite{DKR} we considered
a class of freely acting $\IZ_k$ orbifolds
possessing a global $SU(3)$ symmetry, where
\begin{eqnarray}
r(g_n)={\rm diag}(\omega_k^n,\omega_k^n,\omega_k^n,\omega_k^{-3n})\ ,
\qquad \omega_k = e^{2\pi i/k}
\label{spinSU3}
\end{eqnarray}
and
\begin{eqnarray}
R(g_n) ={\rm diag}(\omega_k^{2n},\omega_k^{2n},\omega_k^{2n},
\omega_k^{-2n},\omega_k^{-2n},\omega_k^{-2n})~~,
\label{vecSU3}
\end{eqnarray}
where $n=1,\ldots, k-1$ labels the twisted sector.
There are two types of operators,
\begin{eqnarray}
O^{\langle i\jbar
\rangle}_n=\Tr(g_n\phi^i\phi^\jbar)-\frac{1}{3}\eta^{i\jbar}\,O_n\ ,
~~~~~~
O_n=\sum_{i=1}^3\Tr(g_n\phi^i\phi^\ibar)
\end{eqnarray}
which form an
octet and a singlet under the $SU(3)$, respectively. Their anomalous dimensions
\begin{eqnarray}
\gamma_{{\bf r},n} = \frac{1}{16\pi^2\,k}\,\delta_{{\bf r},n}
\end{eqnarray}
are found to be
\begin{eqnarray}
\delta_{{\bf 8},n}= 8\lambda \sin^2 (2\pi x)\ ,\qquad
\delta_{{\bf 1},n}= 2\lambda [5+\cos (4\pi x) ]\ ,
\end{eqnarray}
where
\be
x= {n\over k}
\ .
\ee
Introducing
the double-trace tree-level terms
\begin{eqnarray}
\delta S^{\rm tree}_{\rm 2~trace}&=&
\frac{1}{2}\sum_{n=1}^{k-1}\,f_{{\bf 8},n}
O_n^{\langle i\jbar \rangle} O_{-n}^{\langle j\ibar \rangle}
+
\frac{1}{2}\sum_{n=1}^{k-1}\,f_{{\bf 1},n} O_n O_{-n}
\end{eqnarray}
with the symmetry
\begin{eqnarray}
f_{{\bf 8},n} = f_{{\bf 8},k-n}~~~~~~~~ f_{{\bf 1},n} = f_{{\bf 1},k-n}
\end{eqnarray}
leads to the following beta functions:
\begin{eqnarray}
\beta_{8, n}\!\!\!&=&\!\!\!\frac{1}{32\pi^2 k}
\left[(f_{8,n}+2\delta_{8,n})^2
-256 \lambda^2 \left(3+4\cos (2\pi x)\right)
\sin^4 (\pi x )
\right]
\label{octetSU3}
\\
\beta_{1, n}\!\!\!&=&\!\!\!\frac{3}{32\pi^2 k}
\left[\left(f_{1,n}+\frac{2}{3}\delta_{1,n}\right)^2
-\frac{64}{9} \lambda^2 \left(1+2\cos (2\pi x)\right)
\left(4-2\cos (2\pi x) +\cos (4\pi x)\right)
\right]\label{singletSU3}
\end{eqnarray}
The discriminants of both types of beta functions
can become negative. The discriminant of the singlet beta
function is negative for  $1/3<x<2/3$; the negative region of the
octet discriminant, $\cos(2\pi x)<-3/4$,
is completely included in this range.

%


\section{The role of $U(1)$ factors}

Let us consider an orbifold group $\Gamma$ of rank $k=|\Gamma|$.
We start from the ${\cal N}=4$ $SU(kN)$ SYM and get
$U(N)^k/U(1)=SU(N)^k\times U(1)^{k-1}$ as a result. Since the
corresponding gauge group is not simple, we need to introduce
the $U(1)$ coupling constant $e$. \footnote{The renormalization of all
the $U(1)$ couplings is the same in a given orbifold theory since
any node of the corresponding quiver diagram can be mapped into any
other node by some orbifold group element.}
On the
``natural line'' \cite{Fuchs} it is equal to the $SU(N)$ coupling
constant $g$ at tree-level.
Renormalization group flow drives $e$ to smaller values.
There is no inheritance
theorem for $e$ since $U(1)$ does not distinguish the single and double
trace structures. Therefore, $\beta_e$ does not vanish; hence
the orbifold theory is not conformal on the natural line.
Far in the IR $e$ flows to zero and the $U(1)$ factors decouple.

Therefore,
in order to search for a conformal theory we must remove
the $U(1)$ factors and work with the $SU(N)^k$ gauge theory \cite{DKR}.
Let us investigate how this modifies the beta-functions of
other couplings. The background and quantum fields now must
satisfy an additional constraint
\bea
\Tr(g\Phi)=0~~~~
 (\forall)~ g\in \Gamma,~~~g\ne 1~~.
\eea
Therefore the propagator changes from
$\langle \Phi^\alpha_\beta
\Phi^\gamma_\delta \rangle =\displaystyle{ 1\over 2Np^2}
\delta^\alpha_\delta \delta^\gamma_\beta$ to
\be \langle \Phi^\alpha_\beta \Phi^\gamma_\delta \rangle={1  \over
2Np^2}\left[\delta^\alpha_\delta \delta^\gamma_\beta -{1\over
N}\delta^\alpha_\beta \delta^\gamma_\delta \right]\ , \ee
where we
suppressed possible space-time and flavor indices
and did not take orbifold
structure into account because it is not important for our
consideration.

The new term $-{1\over N}\delta^\alpha_\beta \delta^\gamma_\delta$
does not change the leading large $N$ one-loop result either for
single-trace or for double-trace operators.\footnote{
We expect this to be true
also for the higher loops, but have not checked them explicitly.
}
Thus, the one-loop effective potential and the
beta-functions are not modified to leading order in $1/N$
by removing the $U(1)$ factors.
The only modification appears at tree level;
the $U(N)$ potential
\be
\lambda N \Tr\big|[\Phi_\mu,\Phi_\nu]\big|^2=\lambda N \left[
\Tr\big|\sum_i[\Phi_i,\Phi_i^\dagger ]\big|^2 +
4 \Tr\sum_{i\ne j}\big|[\Phi_i,\Phi_j]\big|^2\right]
\ee
is corrected by the addition of the double-trace terms
\be
\label{op}
-\lambda \sum_{g\in \Gamma} \left[
\big|\sum_i\Tr g[\Phi_i,\Phi_i^\dagger ]\big|^2 +4 \sum_{i\ne j}\big|
\Tr g[\Phi_i,\Phi_j]\big|^2\right]
\ee
which arise from integrating out the twisted $U(1)$ auxiliary fields in
the vector and chiral superfields, respectively.
Here $\Phi_i,\ i=1,2,3,$ are the complex scalar fields.

We conclude that the only difference between the $SU(N)^k$ and
$U(N)^k/U(1)$ is the double trace operator that appears
at the tree level. Let us clarify the relation of this fact
to our
previous work \cite{DKR} where the one-loop renormalizability
required us to deform the theory by double-trace operators. Now
we note that
such operators are present already at tree level in the $SU(N)^k$
gauge theories. We will now show that, in all 
supersymmetric orbifolds, the tree-level
coefficients (\ref{op}) are in fact the RG fixed points
of the double-trace couplings.
%
%

\subsection{${\cal N}=2$ supersymmetric orbifolds}

Let us illustrate the discussion above
by considering ${\cal N}=2$ supersymmetric $\IZ_k$
orbifold theories.
We define the action of $\IZ_k$ by
\bea
r=\left(
\begin{array}{cccc}
  e^{im\alpha} & 0 & 0 &  0\\
  0 & e^{-im\alpha} & 0 & 0 \\
  0 & 0 & 1 & 0 \\
  0 & 0 & 0 & 1 \\
\end{array}
\right)~~~~~~~~~~~~~~~~~~~~~~~{\alpha={2\pi\over k}}~~.
\label{actNeq2}
\eea
The scalar field
content of the quiver gauge theory is the following: a complex (adjoint) field
$\Phi=\Phi_3$ and ``bi-fundamental'' fields $Q=\Phi_2$ and
$\tilde{Q}=\Phi_1$ which form an ${\cal N}=2$
hypermultiplet.
These fields  are eigenvectors for $\IZ_k \in SO(6)$
\begin{eqnarray}
\begin{array}{lll}
R(\Phi)=\Phi &~~~~~~ & R({\Phi}^\dagger )={\Phi}^\dagger \cr
R(Q)=Q e^{im\alpha} && R(Q^\dagger )=Q^\dagger  e^{-im\alpha} \cr
R(\tilde{Q})=\tilde{Q}e^{-im\alpha} && 
R(\tilde{Q})^\dagger =\tilde{Q}^\dagger e^{+im\alpha}
\end{array}
~~~~.
\label{explicit}
\end{eqnarray}
%

The divergent part of the double-trace contribution to
the one-loop effective action
%
%
follows from our general formulae:
\bea
&&
\delta S^{\rm 1~loop| 2~tr}=
-{\lambda^2 \Div \over 2 k}
\sum_{n=1}^{k-1} 8(1-\cos(nm\alpha))^2
\left[ \Tr(g_n QQ^\dagger )\Tr(g_n^\dagger QQ^\dagger )
\right. \cr
&&~~~~~
+\Tr(g_n\tilde{Q}\tilde{Q}^\dagger )
\Tr(g_n^\dagger \tilde{Q}\tilde{Q}^\dagger )
- \left.2\Tr(g_n QQ^\dagger )\Tr(g_n^\dagger \tilde{Q}^\dagger
\tilde{Q})+4 \Tr(g_n Q\tilde{Q})\Tr(g_n^\dagger \tilde{Q}^\dagger
Q^\dagger ) \right]
\label{res}\\
&&
=
-{\lambda^2 \Div \over 2 k} \sum_{n=1}^{k-1}
4(1-\cos(nm\alpha))\left[ |\Tr(g_n
[Q,Q^\dagger ])-\Tr(g_n[\tilde{Q}^\dagger ,\tilde{Q}])|^2+4 |\Tr(g_n
[Q,\tilde{Q}])|^2\right] ~~.
\nonumber\eea
Note that the adjoint field $\Phi$ does not appear
in the answer.

In the previous section we observed that the tree-level potential
of the $SU(N)^k$ quiver gauge theory, i.e. the orbifold theory
with the $U(1)$ factors removed, contains the double-trace
operators 
\be {1\over k}\sum_{n=1}^{k-1} {f_n}\left[ |\Tr(g_n
[Q,Q^\dagger ])+\Tr(g_n[\tilde{Q}^\dagger ,\tilde{Q}])|^2+4 |\Tr(g_n
[Q,\tilde{Q}])|^2 \right]\ , \ee 
with coupling  constants
$f_n=-\lambda$. We will now check that precisely for this value
the double-trace beta function vanishes. Indeed, we find \be
\beta_{f_n}=4{(1-\cos (nm\alpha) )\over 16 \pi^2 k}(\lambda
+f_n)^2 ~~. \ee 
Therefore, as expected, there is a fixed line
$f_n=-\lambda$  corresponding to the ${\cal N}=2$ superconformal
$SU(N)^k$ gauge theory.

Let us remark that,
as a consequence of the choice of the orbifold
action (\ref{actNeq2}), the $SU(2)_R$ symmetry of the orbifold theory
is not a subgroup of the $SU(3)$ manifest in ${\cal N}=1$ superspace.
Indeed, the $SU(2)\subset SU(3)$ leaving $\Phi$
invariant and rotating
\be \left(\begin{array}{c}
Q \\
  \tilde{Q} \\
\end{array}\right)
\ee
as a doublet does not commute with the orbifold group
(\ref{explicit}). The $SU(2)_R$-doublets are rather
\be
q_1^i=
\left(\begin{array}{c}
  Q \\
  \tilde{Q}^\dagger  \\
\end{array}\right)~~
~~~~~q_2^i=\epsilon_{ij}(q_1^j)^*=
\left(\begin{array}{c}
  \tilde{Q} \\
  -{Q}^\dagger  \\
\end{array}\right)~~.
\label{rd}
\ee
%

It is straightforward to check that (\ref{res}) is invariant under
$SU(2)$ transformations of (\ref{rd}), being a particular case of the
following manifestly invariant expression:
%
\begin{eqnarray}
&&\!\!\!\!\!\!\!
\Tr(g_n q_1^i q_2^j) Tr(g_n^\dagger q_1^k q_2^l)
(B\epsilon_{ik}\epsilon_{jl}-A\epsilon_{il}\epsilon_{jk})
=  2(A+B) \Tr(g_n QQ^\dagger )\Tr(g_n^\dagger \tilde{Q}^\dagger
\tilde{Q}) \\
&&\vphantom{\Bigg\uparrow}
+A \left(\Tr(g_n QQ^\dagger )\Tr(g_n^\dagger QQ^\dagger )+
\Tr(g_n\tilde{Q}\tilde{Q}^\dagger )
\Tr(g_n^\dagger \tilde{Q}\tilde{Q}^\dagger )\right)
-2B \Tr(g_n Q\tilde{Q})
\Tr(g_n^\dagger \tilde{Q}^\dagger Q^\dagger )~~.\nonumber
\end{eqnarray}
%

\subsection{${\cal N}=1$ supersymmetric orbifolds}

It is not hard to perform a similar analysis for the ${\cal N}=1$
supersymmetric 
$\IZ_k$ orbifolds.
The action of the generator
in the fundamental representation of the R-symmetry group
is
\bea r=\left(
\begin{array}{cccc}
  e^{im_1 \alpha} & 0 & 0 &  0\\
  0 & e^{i m_2 \alpha} & 0 & 0 \\
  0 & 0 & e^{-i(m_1+m_2)\alpha} & 0 \\
  0 & 0 & 0 & 1 \\
\end{array}
\right)~~~~~~~~~~~~~~~~~~~~~~~{\alpha={2\pi \over
k}}\ .\eea
with the action on the 
three complex scalar fields $\Phi_i$ defined in the usual way. 
This action can be easily generalized to
arbitrary abelian orbifolds.

The double-trace terms induced at one-loop level are
\be
\delta S^{\rm 1~loop|2~tr}=
{\lambda^2 \Div \over k}\sum_{n=1}^{k-1}
[3-\cos(nm_1\alpha)-\cos(nm_2\alpha)-\cos(m_1+m_2)n\alpha]
|\sum_i \Tr(g_n[\Phi_i,\Phi_{\ibar}])|^2  \ ,
\ee
where $g_n$ is the $n$-th element of $\IZ_k$.
The tree-level potential of the $SU(N)^k$ theory contains
\be
{1\over k}\sum_{n=1}^{k-1} {f_n}
\Big|\sum_i
\Tr(g_n[\Phi_i,\Phi_{\ibar}])\Big|^2
\label{treeNeq1}
\ee
with $f_n=-\lambda$.
The beta-function for $f_n$ is found to be
\be \label{SUSYbeta}
\beta_{f_n}={1\over 8 \pi^2 k}
[3-\cos(nm_1\alpha)-\cos(nm_2\alpha)-\cos(m_1+m_2)n\alpha]
(\lambda+f_n)^2\ , \ee 
which indeed vanishes for this value of the coupling.
Therefore, as expected,
there is a fixed line of
${\cal N}=1$ superconformal $SU(N)^k$ gauge theories.

This structure of $\beta_{f_n}$ has the following explanation in
the $SU(N)^k\times U(1)^{k-1}$ gauge theory. Since
$\lambda+ f_n\sim e^2 N$, where $e$ is the coupling of the $U(1)$
factors, (\ref{SUSYbeta}) may be rewritten as
\be
\beta_e \sim e^3 N
\ ,
\ee
i.e., the double-trace beta function is related by supersymmetry
to the beta function for the $U(1)$ charge.
Thus, the supersymmetry of the $SU(N)^k\times U(1)^{k-1}$ gauge theory
explains the universal $(\lambda+f_g)^2 $ form of the double-trace
beta function. It follows that the beta functions
for all double-trace operators generated at one-loop level
have vanishing discriminants $D$
in all supersymmetric orbifold gauge theories. 
This fact will be important in comparing the field theory results
with closed string theory.

In the next section we will discuss general abelian orbifold theories
of which the $\IZ_k$ orbifolds discussed here are a special case.
The coupling constants $f_n$ in equation (\ref{treeNeq1})
acquire an additional label $f_n\rightarrow f_{n,i}$, and the existence
of a fixed line will be tested through the signs of the
eigenvalues of a matrix discriminant. In comparing this more
general setup with the present discussion, we will encounter
cases where
\begin{eqnarray}
\beta_{f_O}
=f_O^2+2 \lambda \gamma_O f_O \ ,
\end{eqnarray}
i.e. the operator is not induced by one-loop diagrams.
In these cases $D=\gamma_O^2 > 0$, and there are two solutions,
$f_O=0$ and $f_O=- 2\gamma_O \lambda$. Taking $\gamma_O>0$ we note that
only $f_O=0$ is an IR stable fixed point. Hence, we pick this solution
which preserves supersymmetry.


\section{ General abelian orbifolds}

Now we are ready to consider general abelian orbifolds. All group
elements can be simultaneously brought to a diagonal form
\bea r=
\left(
\begin{array}{cccc}
  e^{2 \pi i x_1} & 0 & 0 &  0\\
  0 & e^{2\pi i x_2} & 0 & 0 \\
  0 & 0 & e^{2\pi ix_3} & 0 \\
  0 & 0 & 0 & e^{-2\pi i(x_1+x_2+x_3)} \\
\end{array}
\right)\ ,
\label{actk}
\eea
perhaps with an additional set of integer labels. For example,
if $\Gamma=\IZ_{k_1}\times \dots \times \IZ_{k_s}$, the exponents
$x_i$ are given by
\begin{eqnarray}
x_i=\sum_{j=1}^s m_{ij}\frac{n_j}{k_j}~~,
\end{eqnarray}
where $m_{ij}$ is a set of integers labeling the group
representation, and $n_j$ label the group elements.
For ease of notation we further introduce the variables
\begin{eqnarray}
y_i=2\pi x_i~~,~~~~ y=y_1+y_2+y_3~~.
\end{eqnarray}


We will show that there is no fixed line unless the
theory is supersymmetric, which happens either
when one of the parameters $x_i$ is
an integer, or when $x_1+ x_2+ x_3$ is an integer.

To reach this conclusion it is sufficient to analyze the
operators $\Tr(g\Phi_i\Phi_{\ibar})\Tr(g^\dagger \Phi_j\Phi_{\jbar})$
where no summation over repeated indices is implied.
Such operators can be constructed from  the spectrum of general
abelian orbifolds. In cases in which some nonabelian global symmetry
survives the orbifold projection, some of the single-trace factors
of the double-trace operators belong to
nontrivial representations of this global symmetry
group, such as $\Tr(g\Phi_i\Phi_{\jbar})\Tr(g^\dagger
\Phi_j\Phi_{\ibar})$. The surviving nonabelian global symmetry
prevents them from mixing with $|\Tr(g\Phi_i\Phi_j)|^2$ when the
latter exist. Thus, the operators $\Tr(g\Phi_i\Phi_{\ibar})
\Tr(g^\dagger \Phi_j\Phi_{\jbar})$ are a universal guide to the
behaviour of orbifold theories.
%

For further convenience we introduce the notation
\be
O^g_i
=
\frac{1}{\cos\frac{1}{2}(y-y_i)}
\Tr(g\{\Phi_i , \Phi_{\ibar}\})
=
\frac{1}{\cos\frac{1}{2}(y-y_i)}
\Tr(g\Phi_i  \Phi_{\ibar}+g \Phi_{\ibar}\Phi_i) \ .
\ee
The prefator in the definition above is necessary to make the operator
$O^g_i$ well-defined for the orbifold group elements which anticommute
with at least one of the scalar fields.

In terms of $O_i^g$ the 1-loop effective potential could be written as
\be
\delta S^{1~{\rm loop}| 2~{\rm tr}}
={\lambda^2 \Div\over 16\pi^2 k}\sum_g \sum_{i,j}
~\cos\frac{1}{2}(y-y_i)\cos\frac{1}{2}(y-y_j)~
h^g_{ij}
O^g_i O^{g^\dagger }_j~~.
\label{eff}
\ee
The reality of the effective action
requires that the matrix $h^g_{ij}$
be hermitian
$(h^g_{ij})^*=h^{g^\dagger }_{ij}=h^g_{ji}$; It turns out however
that it is actually real and symmetric.
All its elements can be obtained by a suitable relabeling of the entries
$h^g_{11}$ and $h^g_{12}$:

\bea
\cos^2({y_2+y_3 \over 2})
h^g_{11}&=&-
8\cos(\frac{y_1+y_2}{2})\cos(\frac{y_2+y_3}{2})\cos(\frac{y_3+y_1}{2})
+ 2(1-\cos(y_2+y_3))
 \\
&+&\!\big(1+\cos^2({y_2+y_3\over 2})\big)
(1+\cos(y_1+y_2)+\cos(y_2+y_3)+\cos(y_3+y_1))
\nonumber
\eea
%
%
\bea
&&\cos({y_2+y_3\over 2})\cos({y_1+y_3\over 2}) h^g_{12}=
-4\cos(\frac{y_1+y_2}{2})  \\
&& ~~~~~~
+\cos({y_2+y_3\over 2})\cos({y_1+y_3\over 2}) \big[
5+ \cos({y_1+y_2})
- \cos(y_2+y_3)-
\cos(y_1+y_3) \big] ~~.
\nonumber
\eea

The effective action (\ref{eff}) implies that the
deformation (\ref{def}) involves a hermitian matrix
of tree-level couplings $f^g_{ij}$
\be
\delta S_{tree}= {1\over 2 k}\sum_g{f^g_{ij} }O^g_i
O^{g^\dagger }_j ~~.
\ee
The explicit form of the
corresponding beta-function is rather involved,
\bea
\beta_{f^g_{ij}}&=&{1\over 16\pi^2 k}\left( 2\lambda^2
\left[\cos\frac{1}{2}(y-y_i)\cos\frac{1}{2}(y-y_j) h^g_{ij}\right]
+2\lambda f^g_{ij} (2-\cos(y-y_i)-\cos(y-y_j)) \right.
\nonumber \\
&&\left. +2 \lambda f^g_{ik} \cos({y-y_k\over 2})\cos({y-y_j\over 2})
+2\lambda f^g_{kj}\cos({y-y_i\over 2})\cos({y-y_k\over 2})
+ f^g_{ik}f^{g}_{kj}         \right)\ ,
\eea
%
but can be rearranged in a convenient form
\be
\tilde{\beta}_{f^g_{ij}}=
(\tilde{f}^g_{ik}+A^g_{ik})(\tilde{f}^g_{kj}+A^g_{kj})-4\lambda^2M^g_{ij}
\ee
exposing the matrix analog $M^g_{ij}$ of the discriminant $D$. Here
\be
A^g_{ij}=2\lambda\left(\delta_{ij}(1-\cos({y-y_i}))+
\cos({y-y_i\over 2})\cos({y-y_j\over 2}) \right)\ ,
\ee
and the explicit form of the hermitian matrix $M^g_{ij}$ is
\begin{eqnarray}
\label{matrix}
M^g_{ij}&=&-{1\over 2}\cos({y-y_i \over 2}) \cos({y-y_j \over
2})h^g_{ij}+
{1\over 4\lambda^2 }A^g_{ik}A^g_{kj}\ ,\\
M^g_{ii}&=&1-\frac{1}{2}\cos{(y-y_i)}-
\frac{1}{2} \sum_{k=1}^3\cos{(y-y_k)}
+4\prod_{k=1}^3 \cos({y-y_k \over 2})\ ,\\
M^g_{ij}&=&\cos({y-y_i \over 2})\cos({y-y_j \over 2})
+2 \cos({y_i + y_j \over 2}) \Big|_{i\ne j}~~.
\end{eqnarray}
This matrix is obviously invariant under $y_i\rightarrow 2\pi-y_i$ for all $i=1,2,3$:
this is a manifestation of the symmetry under $g \rightarrow g^{-1}$.
Another symmetry is related to the choice of basis on the maximal torus  of $SU(4)$.
The change $y_3\rightarrow -y_1+y_2+y_3$ combined with
interchange of indices $1$ and $2$ ($1 \leftrightarrow 2$)
leaves the  matrix $M_{ij}$ invariant.

If we can find an orbifold group element
$g$ such that at least one eigenvalue of $M$ is negative,
then at least one beta function does not vanish for any real
value of the corresponding tree-level coupling and consequently
conformal invariance is broken.
%
%
Therefore, the smallest eigenvalue
of $M_{ij}$ as a function of $x_i$ is a useful quantity which
controls the existence of a fixed line.

Now we find the region inside the unit cube $0 \leq x_i<1$, i.e.
$0\leq y_i<2\pi$, where the smallest eigenvalue is negative. Let us
start from the points where supersymmetry is preserved. They are
the edges of the cube $x_i=0$ or the planes  $x_1+x_2+x_3=1,2$.
The fixed line exists in these cases, as was shown in section 3.
For example, if $y_3=-y_1-y_2$ we have
\bea
M_{ij}&=&{1\over 4\lambda^2}(\tilde{f}+A)^2~~,\\
{1\over 2\lambda}(\tilde{f}+A)&=&\left(%
\begin{array}{ccc}
  1 & \cos(y_3+{y_1+y_2\over 2}) & \cos(y_2+{y_1+y_3\over 2}) \\
 \cos(y_3+{y_1+y_2\over 2}) & 1 & \cos(y_1+{y_2+y_3\over 2}) \\
 \cos(y_2+{y_1+y_3\over 2}) & \cos(y_1+ {y_2+y_3\over 2}) & 1 \\
\end{array}%
\right)  \label{fm} \\
y_3&=&-y_1-y_2
\eea
and therefore for any choice of $y_i$
its smallest eigenvalue is equal to zero.
\footnote{ ${1\over 2\lambda}(\tilde{f}+A)$ has the form (\ref{fm})
only for $y_3=-y_1-y_2$. Changing the choice of supersymmetry
constraint requires appropriate reshuffling of $y_{1, 2, 3}$ in
(\ref{fm}).
}

A further analysis shows that these are the only hypersurfaces where
the smallest eigenvalue of $M_{ij}$
is equal to zero. As a result, the cube $0 \leq x_i<1$ is divided into
three regions: $0\leq x_1+x_2+x_3\leq 1$,
$1< x_1+x_2+x_3< 2$ and $2\leq x_1+x_2+x_3\leq 3$. In the
first and the third regions, which are related by the $x_i \rightarrow 1-x_i$
symmetry, all eigenvalues of $M$ are non-negative. In the ``middle'' region $1< x_1+x_2+x_3< 2$
at least one eigenvalue of $M$ is negative: this is the unstable
region. This structure is shown in Figure 1.

\section{Closed String Zero-Point Energy}

To calculate the spectrum of twisted closed strings on $\IR^{3,1}
\times \IR^6/\Gamma$, it is efficient to use the light-cone gauge
Green-Schwarz methods. We refer, e.g., to
\cite{KNS} for an application to a simple $\IZ_4$ orbifold
which we now generalize to arbitrary
abelian orbifolds.

Let us consider first $\Gamma=\IZ_k$. Then, the equation (\ref{actk})
gives the action
of $\IZ_k$ in the fundamental representation of $SU(4)$, while
the action in the vector representation of $SO(6)$ is
\begin{eqnarray}
R(g_n) ={\rm diag}(\omega_k^{n(m_1+m_2)},\omega_k^{n(m_1+m_3)},
\omega_k^{n(m_2+m_3)},
\omega_k^{-n(m_1+m_2)},\omega_k^{-n(m_1+m_3)},\omega_k^{-n(m_2+m_3)})
\ ,
\end{eqnarray}
where $\omega_k=e^{2\pi i/k}$.

In the light-cone Green-Schwarz approach one finds four complex world
sheet bosonic coordinates $X^l$ and four complex fermionic coordinates
$b^l$. In the $n$-twisted sector we find that the fermionic
cordinates have boundary conditions
\begin{equation}
b^l (\sigma + 2\pi) = e^{2\pi i n m_l/k} b^l (\sigma)
\ ,
\end{equation}
and the bosonic coordinates
\begin{eqnarray}
&&X^1 (\sigma + 2\pi) = X^1 (\sigma)
\ , \nonumber \\
&&X^2 (\sigma + 2\pi) = e^{2\pi i n (m_1 + m_2)/k} X^2 (\sigma)
\ , \nonumber \\
&&X^3 (\sigma + 2\pi) = e^{2\pi i n (m_1 + m_3)/k} X^3 (\sigma)
\ , \nonumber \\
&&X^4 (\sigma + 2\pi) = e^{2\pi i n (m_2 + m_3)/k} X^4 (\sigma)
\ .
\end{eqnarray}
These equations can be trivially extended to general abelian
orbifolds (\ref{actk}) by making the replacements
$n m_l/k\rightarrow x_l$, etc.
\footnote{It is worth pointing out that
the equation (\ref{actk}) covers all possible orbifold
actions. Similarly to the $\IZ_4$ case discussed in \cite{KNS},
orbifolds for which the values
$x_i=1/2$ are allowed can be thought of as orbifolds of the
type 0 theory rather than of the type $II$ theory. One may
translate them to the type II theory by multiplying the fermion
action by $(-)^{\bf F}$ \cite{ADPS}. This simply amounts to the
redefinition $x_i\rightarrow x_i+1/2$ which leads to a reshufling of
the fermion charges. This shift is inconsequential for the orbifold
action in the vector representation.}

The zero-point energy due to a complex boson with boundary condition
$ X (\sigma + 2\pi) = e^{2\pi i \theta} X (\sigma) $ is
\begin{equation}
{1\over 24} - {( 2 \{ \theta \}  -1 )^2\over 8} \ ,
\end{equation}
where $\{ \theta \}$ is the fractional part of $\theta$,
which ranges from 0 to 1.
For a fermion we get the opposite of this quantity.
Using the variables $x_i
$, $i=1,2,3$
introduced in the previous section, we find that
the energy of the ground state is given by
\begin{eqnarray}
 8 E_0 (x_1,x_2,x_3)&=& -1 - ( 2 \{ x_1+ x_2 \}  -1 )^2 - ( 2 \{ x_1+
x_3 \}  -1 )^2 - ( 2 \{ x_2+ x_3 \}  -1 )^2 \nonumber \\
&&+ ( 2 \{ x_1+ x_2+ x_3 \}  -1 )^2 + \sum_{i=1}^3
( 2 \{ x_i  \}  -1 )^2\ .
\label{zeropt}
\end{eqnarray}
Since $E_0 (x_1,x_2,x_3)$ is periodic in each of the variables with period $1$,
it is sufficient to determine it inside the unit cube $0\leq x_i < 1$.
Just as the matrix $M_{ij}$, $E_0$
also has a discrete symmetry
\be \label{discsym}
E_0 (1-x_1,1- x_2, 1-x_3)=
E_0 (x_1,x_2,x_3)\ .
\ee

Let us divide the unit cube into three regions by the planes
$x_1+ x_2+ x_3= 1$ and $x_1+ x_2+ x_3= 2$, as in Figure 1.
In the lower region
$0 \leq x_1+ x_2+ x_3 \leq 1$, we also have
$0 \leq x_i+ x_j\leq 1$, so that all $\{\}$ symbols may be removed
from (\ref{zeropt}). Then it is easy to see that $E_0$ vanishes
identically in the lower region. By the symmetry (\ref{discsym})
this is also true in the upper region.

In the region between the planes, $1 < x_1+ x_2+ x_3 < 2$,
the function has a different form depending on whether
$x_i+x_j$ exceeds unity, but it is not hard to see that $E_0$
is negative in the entire middle region.
Thus, only the middle region corresponds to a tachyonic instability.

\begin{figure}[ht]
\begin{center}
\epsfig{file=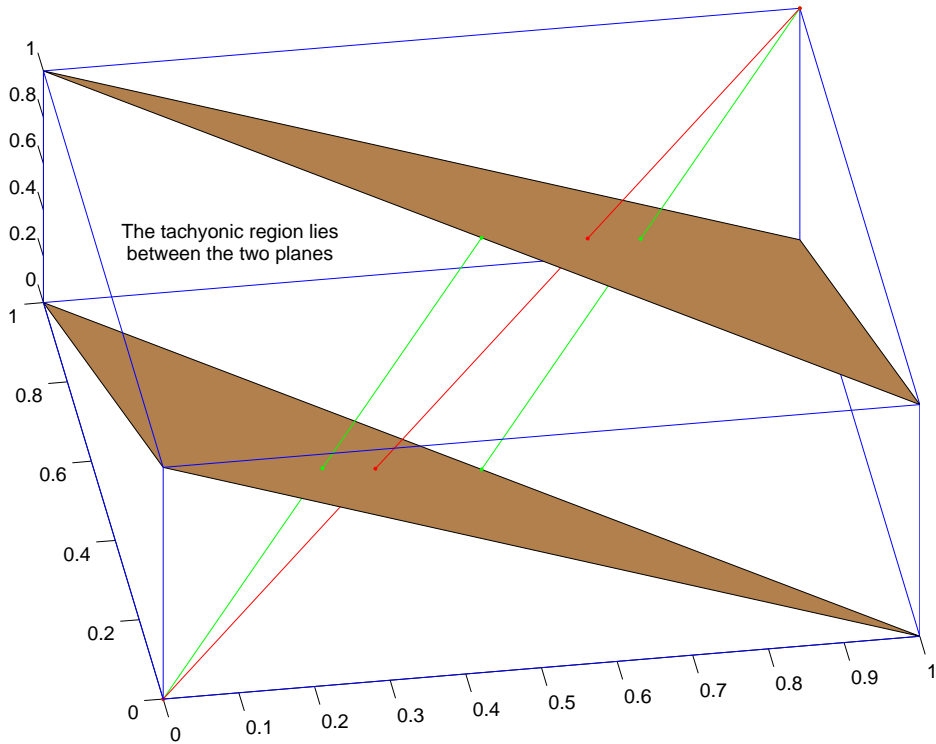,height=10cm}
\caption{Stability/instability regions for general abelian 
orbifolds \label{fig} }
\end{center}
\end{figure}

\section{Comparison}

In the last two sections we showed that
the region of instability of the one-loop orbifold
gauge theory coincides with that of the free closed
superstrings on $\IR^{3,1}\times \IR^6/\Gamma$.
This is summarized in Figure \ref{fig}, the unstable region
being contained between the two shaded planes. The three axes
correspond to the $x_i$ with $i=1, 2, 3.$
A point in this region corresponds both to a gauge theory operator
with nonvanising beta function as well as to a closed string
tachyonic ground state.
%
%
Here we further discuss this surprising correspondence
and illustrate it with some simple examples.

We will start with supersymmetric theories. The orbifold
preserves supersymmetry if and only if all the double-trace
operators
correspond to points on the hypersurfaces
$x_1+x_2+x_3=0~ mod~ 1$. In these cases the string zero-point
energy is equal to zero and also equal to the smallest
eigenvalue of the matrix $M$.
A simple example of such a theory is the $SU(3)$ symmetric
$\IZ_3$ orbifold.
As discussed in \cite{DKR}, this theory has an untwisted sector,
$n=0,x_i=0$, and two twisted sectors, $n=1,x_i=1/3$ and $n=2,x_i=2/3$.
All these points (red dots) lie on the ``supersymmetric'' hypersurfaces
$x_1+x_2+x_3=0~ mod~ 1$ as shown
in Figure \ref{fig}.

Now let us discuss the class of $\IZ_k$ orbifolds with $SU(3)$ global
symmetry. This class of theories was
discussed extensively in \cite{DKR} and corresponds to $m_1=m_2=m_3$,
i.e. to moving along
the line $x_1=x_2=x_3=x$ which is shown in red in Figure 1.
In this case $E_0=0$ for $0\leq x \leq 1/3 $, and
\be \label{SU3ground}
E_0 (x) =1 - 3x \ ,\qquad
1/3 \leq x \leq 1/2 \ .
\ee
Thus, the twisted sector ground states with small
charges, $n< k/3$, are not tachyonic, but for
$k/3 < n < 2k/3$ they are.
Comparing this with the plot of the
singlet discriminant $D$ in the field theory calculation of
\cite{DKR}, we observe a remarkable
agreement: a transition from stability to instability at $x=1/3$ and
back to stability at $x=2/3$.
This result was already derived in \cite{DKR}, but
let us trace here how it appears from the matrix (\ref{matrix}).
In this case $M_{ij}$ may be simplified to
\bea
M_{ii}=-4\cos^2(2\pi x)+3+4\cos^3(2\pi x) \ , %
~~~~~~~~
M_{i\neq j}=\cos(2\pi x ) (\cos(2\pi x) + 2)~~.
\eea
The eigenvalue which becomes negative first as $x$ is increased,
\be \left(1+2\cos (2\pi x)\right)
\left(4-2\cos (2\pi x) +\cos (4\pi x)\right)
\ ,\ee
coincides up to an irrelevant numerical factor with the discriminant
for the singlet beta-function (cf. equation (\ref{singletSU3})).
Its sign changes from positive to negative
values at $x=1/3$, and back to positive at $x=2/3$, in perfect
agreement with the closed string zero-point energy (cf. the solid 
line portions of the two curves in Figure 2).

\begin{figure}[ht]
\begin{center}
\epsfig{file=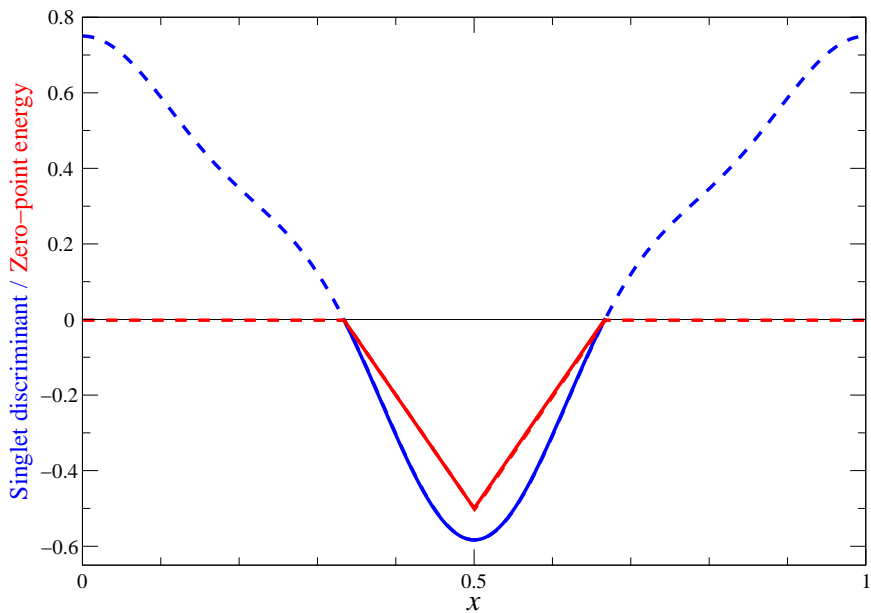,height=8cm}
\caption{
The (rescaled) singlet discriminant and the closed string
zero-point energy. Both are negative in the same range of $x$.
}
\end{center}
\end{figure}

The other two eigenvalues coincide and are proportional to the
discriminant of the octet beta function (\ref{octetSU3}):
\footnote{Although we reproduce
the result of \cite{DKR}, note that here we discuss only
the ``diagonal'' operators $\Tr(g\Phi_i\Phi_{\ibar})$ while
in \cite{DKR} more general operators were considered.
As discussed at the beginning of secton 4, the apparently missing
operators can be obtained from the ones we considered here through
the action of
$SU(3)$, which for this choice of orbifold
survives as nonabelian global symmetry.
}
\be
\left (3+4\cos(2\pi x)\right ) \left (1+ \cos(2\pi x) \right )^2 \ ,
\ee
which crosses from positive to negative values at $\cos(2\pi x)=-3/4$, i.e.
at $x\approx 0.385$. Since this is greater than $1/3$, the octet instability
indeed
sets in after the initial singlet instability. We suggest that the octet
instability is associated with an $SU(3)$ octet excited closed string
state becoming tachyonic.
Let us consider the region $1/3 < x < 1/2$ where
the ground state energy is $(1-3 x)$.
For each of the three twisted complex worldsheet fields $X^i$, the lowest
oscillator excitation energy in this case is
$ (1- 2 x) $.
So, the excited closed strings obtained by acting with one of these
oscillators on the left and one on the right,
\be \alpha^i_{2x-1} \tilde\alpha^{\jbar}_{2x-1} |0\rangle
\ ,
\ee
are tachyonic for
$ 2 - 5x  < 0$,
i.e. for $x> 2/5$. Numerically, this is close to the gauge theory
value $x\approx 0.385$. Perhaps the lack of exact agreement can be
attributed to the fact that the singlet tachyon is already condensed,
which has not been taken into account in our 
calculations.

Another simple example, which has $SU(2)$ symmetry, is
$m_3=2m_2=2m_1$.
This corresponds to the line
$x_1=x_2=x$, $x_3= 2x$, which is shown in blue
in Figure 1. In this case the transition from stability to
instability takes place at $x=1/4$, and $x=1/2$ is an isolated
supersymmetric point that is stable.
This behavior is shown in the Figure 3.

\begin{figure}[ht]
\begin{center}
\epsfig{file=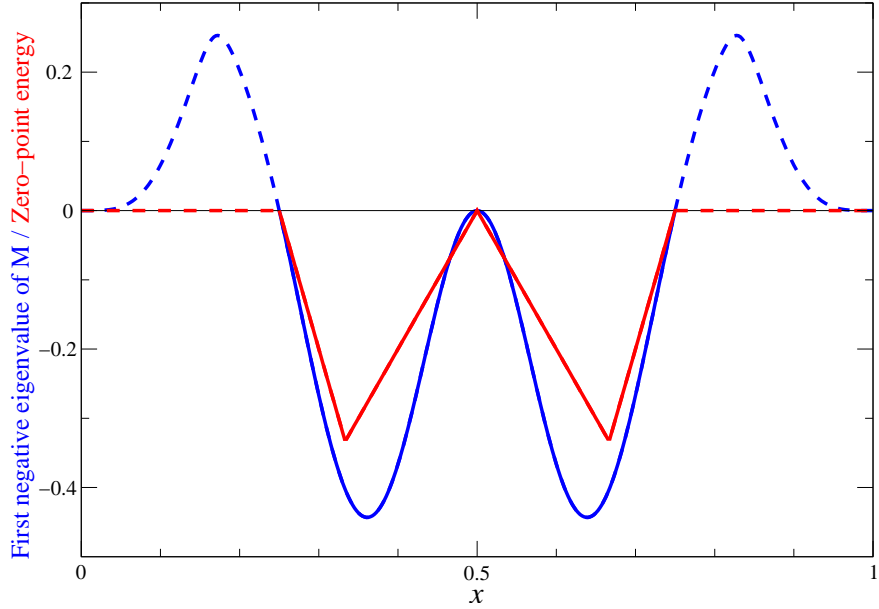,height=8cm}
\caption{
The (rescaled) eigenvalue ${\cal M}_{1}^d$  
and the closed string zero-point energy, plotted
as a function of $x$. Both are negative (the solid
line portions of the two curves) in the 
same range of $x$.}
\end{center}
\end{figure}

The first eivenvalue of the matrix $M$ to become negative
is ($z=\cos(2 \pi x)$)
\begin{eqnarray}
{\cal M}_{1}^d=\textstyle{\frac{1}{2}}\left(
4 + z (12 - 15 z - 8 z^2 + 16 z^3) - (2 + z)
\sqrt{4 - 12 z + z^2 + 16 z^3}\right)~~;
\label{firstn}
\end{eqnarray}
%
it is negative for $1/4 < x < 1/2$, in agreement with the behavior
of the zero-point energy. The ranges of $x$ for which the other two
eigenvalues become negative 
\begin{eqnarray}
{\cal M}^d_2&=&\textstyle{\frac{1}{2}}\left(
4 + z (12 - 15 z - 8 z^2 + 16 z^3) + (2 + z)
\sqrt{4 - 12 z + z^2 + 16 z^3}\right)\cr
{\cal M}_{3}^d&=& (1 - z)^2 (1 + 8 z + 8  z^2)
\end{eqnarray}
are completely contained in $1/4 < x < 1/2$. This suggests that the
subleading instabilities associated with the higher
eigenvalues crossing zero are again dual to excited closed strings
becoming tachyonic. This would be an interesting comparison to pursue
in the future.

\begin{figure}[ht]
\begin{center}
\epsfig{file=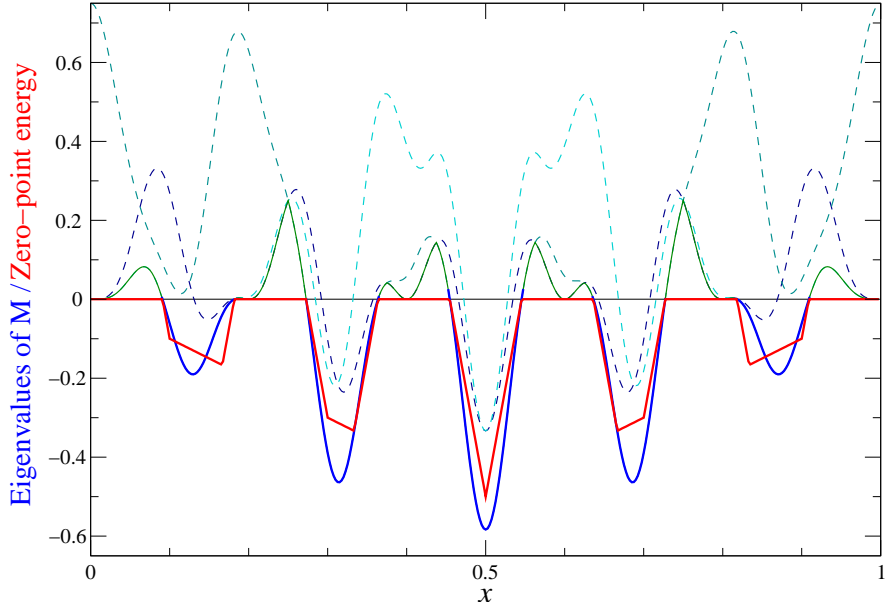,height=8cm}
\caption{
The (rescaled) eigenvalues of the discriminant matrix, the smallest
eigenvalue (continuous line) and the closed string
zero-point energy as a function of normalized twisted sector charge $x$
for an $SU(2)$-symmetric orbifold. The crossing of
eigenvalues as a function of charge is cricial for the successful
match of gauge theory and string theory instabilities.
\label{fig:cross}}
\end{center}
\end{figure}

Generically, the eigenvalues of the disciminant matrix cross as
functions of the twisted sector charge. This effect is in fact crucial
for the agreement of the gauge theory ans string theory instabilities.
An example corresponding to the $SU(2)$-symmetric orbifold
with weights $5m_1=m_2=m_3$ is shown in Figure \ref{fig:cross}.
It is easy to identify on the figure the points where the minimum
of the three eigenvalues (the continuous line) jumps from one
eigenvalue to another (dashed lines). The regions in which it is 
negative belongs to different eigenvalues depending on the twisted
sector charge and agrees with the closed string prediction for the
existence of instabilities.

It was derived earlier  \cite{DKR} that a non-supersymmetric
orbifold gauge
theory with a fixed point, i.e. with $x_3=1-x_2$, has no stable
twisted sectors.
We can easily identify this situation in Figure 1.
For $0< x_1 <1$ and $x_3=1-x_2$,  we always have
$1<x_1+x_2+x_3<2$.

\subsection{A no-go theorem}

Now we are ready to prove a ``no-go theorem'' for non-supersymmetric
conformal orbifolds of ${\cal N}=4$ SYM.
Let us assume that we have a nonsupersymmetric orbifold 
and focus on an element $g$ which, after being diagonalized, becomes
$r(g)={\rm diag}(e^{2\pi i x_1},e^{2\pi i x_2},e^{2\pi i x_3},e^{-2\pi
i X})$ with $X=x_1+x_2+x_3~mod~1$,  
$x_i\neq 0$ for all $i$ and $x_1+x_2+x_3 \neq 0 ~mod~ 1$. In the
appendix we argue that, in agreement with the intuition, such an 
element always exists.
As we have seem in section 4, if $1<x_1+x_2+x_3<2$ this sector 
is unstable. 
The remaining two cases, $0<x_1+x_2+x_3 <1$ and $2 < x_1+x_2+x_3< 3$,
which are related by the transformation $g \rightarrow g^{-1}$ require
us to consider the subgroup of the orbifold group
$\Gamma$ generated by $g$. Since the orbifold group is assumed to be
finite, this subgroup is isomorphic to $\IZ_k$ with some finite $k$.
If the generator of this $\IZ_k$ has all nonvanishing $x_i={m_i/k}$ and
$0<x_1+x_2+x_3<1$, the element $g^n$ with a sufficiently large
$n$ is such that $1<n(x_1+x_2+x_3)<2$ and thus potentially yields 
an unstable sector.

A subtlety which requires a more detailed discussion stems from the
fact that sometimes $nx_i$ or $nX$ are integers and the corresponding
group element preserves supersymmetry. Let us recall the
partition of the domain of $(x_1,x_2,x_3)$ introduced in section 4.
%
%
%
%
%
%
%
%
%
%
%
We called the part of the cube
$x_1+x_2+x_3\leq 1$ (with all edges included i.e. with planes $x_1=0$,
$x_2=0$, $x_3=0$, $X=1$)
the ``first'' region, the part $2\leq x_1+x_2+x_3<3$
(with only  one edge $x_1+x_2+x_3$=2 included) the ``second'' region
and the area $1<x_1+x_2+x_3<2$ (without any edges included)
the ``negative'' region.

The element $g$ described above is just a point in the ``first'' region.
Now let us consider an arbitrary point $g'$ in the ``first'' region.
We assume it is also from our $\IZ_k$. The multiplicative group
operation translates into an additive operation at the level of the
parameters $x_i^g$: $x_i^{g'+g}=x_i^{g'}+x_i^g$.
The crucial observation is that, due to the fact that $x_i<1$,
the point $g'+g$ cannot lie in the ``second'' region for any $g$
and $g'$ from the ``first'' region. The point $g+g'$ could be in
the ``first'' or ``negative'' region only.
%
%

This observation is actually enough to prove the ``no-go theorem.''
We start with a point in the ``first'' region, corresponding to $g$,
but must finish in the ``second'' region, corresponding to
$g^{k-1}= g^{-1}$. Since, as we argued above, we can not jump directly
from the ``first'' to the ``second'' without being in the ``negative''
region at least once, there is at least one group element which
corresponds to a point in the ``negative'' region and the
corresponding  sector is unstable.
It therefore follows that
no non-supersymmetric orbifold of ${\cal N}=4$ SYM
is conformal in the one-loop treatment.



\section{Discussion}

In motivating the AdS/CFT correspondence, it is customary to begin
with the 3-brane metric
\be
h^{-1/2} dx^2 + h^{1/2} (dr^2 + r^2 d\Omega^2)
\ ,
\ee
containing the warp factor $h(r)= 1+ L^4/r^4$, and then taking
a decoupling limit where $r^4 \ll L^4$ so that the constant term
in the warp factor is neglected.
Since $L^4 = a \lambda$, where $a$ is a proportionality constant, the dual
of a weakly coupled gauge theory is a highly curved space $AdS_5\times
S^5/\Gamma$
of small radius $L$. The spectrum of closed strings in such a background
is difficult to compute reliably.

But let us imagine taking $\lambda$ to be small before the decoupling
limit is taken. Then, the back-reaction of the stack of $N$ D3-branes
is small, and we can expand $h^{1/2} (r) = 1+ {a\lambda\over 2 r^4} +
\ldots$
in the metric. The leading term in this expansion corresponds to a closed
string background $\IR^{3,1}\times \IR^6/\Gamma$ where the spectrum of
closed strings can be studied with ease. Of course, a suggestion that
closed string theory on this background is somehow dual to the weakly
coupled orbifold gauge theory seems impossible in general.
But perhaps the twisted sector states, localized near the tip of the cone,
do incorporate some features of the weakly coupled gauge theory (this
is to some extent suggested by the research on localized tachyons
\cite{ADPS,HKMM,Martinec,HEMT}). What we have found in this paper is
that the stability/instability of this background is actually in
one-to-one correspondence with the stability/instability of the
orbifold gauge theory on $N$ D3-branes placed at the tip of the cone.
What are the reasons for this agreement? One possibility is that the
stability is a ``topological'' observable that is not affected by
taking various limits, i.e. the stability can be studied reliably even
before the decoupling limit is taken. Our results indeed suggest
that the standard arguments for the decoupling of low-energy gauge theory
on $N$ D3-branes at the tip of the cone
from closed string modes fail when the closed string spectrum
contains tachyons, either localized or bulk.
Clearly, we need a better understanding of the effect of twisted closed
string modes on the low-energy gauge theory.

It would also be interesting to use our new results
to study the closed string tachyon condensation
in more detail. At least for
the non-freely acting orbifolds, its gauge theory dual \cite{Adams,TZ} is
the Coleman-Weinberg
mechanism \cite{Coleman}
where single-trace operators charged under $\Gamma$ acquire
expectation values, thus breaking the quantum symmetry of the orbifold
gauge theory. It would be interesting to study whether a similar mechanism
is at work also for the orbifolds that are
freely acting on the 5-sphere; for such orbifolds the tachyons are
localized at the tip of the cone.
Finding relations with existing research on localized
tachyon condensation (see, for example, \cite{ADPS,HKMM,Martinec,HEMT}) may
lead to progress in this direction, as well as to possible insight into
the phase transition and subsequent tachyon condensation in the small
radius regime of $AdS_5\times S^5/\Gamma$.
The analysis of localized tachyon condensation on $\IC^3/\IZ_n$ led
to the conclusion that in type II theories the singularity is
completely resolved \cite{MNPL}.
Turning on the tachyon vev was identified in a probe-brane theory
with turning on twisted Fayet-Iliopoulos terms. Given the surprising
agreement we found, it is possible that turning on a tachyon vev below
the phase transition radius has a similar interpretation. 
In fact,
the double-trace deformations we needed to
introduce at tree level for general abelian orbifold theories
off the natural line arise from integrating out the auxiliary fields
descending from the ${\cal N}=4$ $D$ auxiliary 
fields. 
As in the probe-brane analysis,
turning on twisted Fayet-Iliopoulos terms corresponds to further
deforming the action by the operators $O^g$.
In the gauge theory, this is actually a simple consequence of
a twisted operator developing a vev:
when a double-trace operator
$O^g O^{g^\dagger}$ is present in the action and $ O^{g^\dagger}$
condenses, this effectively deforms the action by
a single-trace operator $O^g$.
In the presence of such a deformation the features of the RG flow
are quite different compared to the $\Gamma$-symmetric case.
In particular,
below the scale of the condensate, the gauge coupling constant
acquires a nontrivial beta function.
Matching the IR properties of
such a deformed gauge theory with the localized tachyon
condensation is an interesting problem.

Another interesting direction to pursue is the identification of
``subleading instabilities'' in the gauge theory, which are due to higher
eigenvalues of $M$ crossing zero, with excited closed strings becoming tachyonic.
As we showed for the class of $SU(3)$ symmetric orbifolds, the onset
of the octet instability as a function of $x$ is numerically very close
to the value of $x$ where
an excited free closed string becomes tachyonic. That it is not exactly
the same could be due to the fact that a singlet tachyon is already
condensed, which has not been taken into account in our calculations.
Perhaps this issue could be studied in more detail.

\section*{Acknowledgments}
We are grateful to A. Parnachev and E. Weinberg for useful discussions.
A.~D. and R.~R. would like to thank the Third Simons Workshop in
Mathematics and Physics, where a part of this work was done.
Some of I.~R.~K.'s work on this project took place
at the Aspen Center for Physics, which he thanks for hospitality.
The research of A.~D. is
supported in part by grant RFBR 04-02-16538
and by the National Science Foundation Grant No.~PHY-0243680.
The research of I.~R.~K. is
supported in part by the National Science Foundation Grant
No.~PHY-0243680.
The research of R.~R. was supported in part by funds provided by the
U.S.D.O.E.
under co-operative research agreement
DE-FC02-91ER40671.
Any opinions, findings, and conclusions or recommendations expressed in
this material are those of the authors and do not necessarily reflect
the views of the National Science Foundation.

\newpage


\appendix

\section{$\!\!\!\!\!\!.$~A nonsupersymmetric element of
nonsupersymmetric orbifold}

To complete the proof in section 6.1 we now argue that,  
if the orbifold group breaks supersymmetry,
there always exists an element $g$ of $\Gamma$ which
acts nontrivially on all spinors. This is obviously true if the
orbifold  group is abelian: since the orbifold breaks all
supersymmetries it follows that, after diagonalizing all group
elements, different elements of $\Gamma$ preserve different spinors.
Then, the desired 
group element $g$ is simply the product of sufficiently many
elements of $\Gamma$ preserving spinors. If each group generator
preserves one spinor it is typically sufficient to simply multiply two
such generators. Special care must be taken when some spinor has
opposite charges with respect to the two generators:
\begin{eqnarray}
g_1|\psi\rangle=e^{ix}|\psi\rangle
~~~~~~
g_2|\psi\rangle=e^{-ix}|\psi\rangle
~~~~\rightarrow~~~~
g_1g_2|\psi\rangle=|\psi\rangle
~~.
\end{eqnarray}
If this is the case it suffices to raise one of the generators to some
power different from unity. It is not hard to see that this leads to the
desired result. Indeed, given our assumption we have to analyze only two
charges for each generator; we denote them as the ordered pairs
$(x^{(1)}_2, x^{(1)}_3)$ and $(x^{(2)}_2, x^{(2)}_3)$.
If $x_2^{(1)}=-x_2^{(2)}$ it is necessary to find an
integer $n$ such that $n(x^{(1)}_2, x^{(1)}_3)+
(-x^{(1)}_2, x^{(2)}_3)\ne (0, 0)~mod~1$. Any
integer different from $-[x_3^{(2)}/x_3^{(1)}]$
has the desired effect.

The situation is less transparent if the orbifold group is nonabelian,
as the generators cannot be simultaneously diagonalized. The goal is
to show that, starting from a generator $g_1$ preserving a spinor
$|\psi\rangle$, it is possible to find a group element $g_2$
preserving a different spinor such that for some $n\in\IZ$, $g_2^n
g_1$ preserves no spinor. Without loss of generality we can pick $g_1$
to be diagonal. We can also assume that $g_2$ has one unit eigenvalue
(if it did not then $g_2$ is the group element we are looking for) and
that the spinor preserved by $g_2$ is different from the spinor
preserved by $g_1$ (if no such $g_2$ exists then $\Gamma$ preserves
supersymmetry). The condition that $g_2$ belongs to a
subgroup of $SU(4)$ implies then that only two of its eigenvalues are
independent. We can always write $g_2$ as $g_2=vg_2^dv^\dagger$, where
$v$ is some unitary matrix and $g_2^d$ is diagonal.

The condition that the product $g_2 g_1$ has one unit eigenvalue
\begin{eqnarray}
\det(g_2 g_1-\id )=0
\label{unitev}
\end{eqnarray}
can be thought of as giving an additional relation between the
eigenvalues od $g_2$. We distinguish two cases.

\noindent
1)
If no orbifold group element (not necessarily a generator, and except
$g_1^k$ for all $k$) satisfies this relation then we can pick any one of
them as $g_2$.

\noindent
2)
If there exists one group element such that the equation above is
satisfied, then
\begin{eqnarray}
\det(g_2^n g_1-\id )=0
\label{unitevn}
\end{eqnarray}
is not satisfied for some $n$. Indeed, since for any $n$ $g_2^n$ is
diagonalized by the same matrix $v$ as $g_2$, we can think of this
equation as providing additional constraints on the eigenvalues of
$g_2$. Since under our assumptions $g_2$ has only two independent
eigenvalues, this equation can be satisfied only for one value of
$n\ne 1$. If this value exists it is sufficient to pick a value of $n$
different from it.

Thus, if $g_2$ generates a group larger than $\IZ_3$, it is always
possible to find an element in $\Gamma$ which does not leave any
spinor invariant.  For subgroups of $SU(4)$ for which each element
preserving some spinor generates a $\IZ_k$ with $k\le 3$ a separate
analysis is needed. Luckily there are few such groups \cite{HaHe} and
it  is straightforward though quite tedious to analyze them one by
one. Even in these cases it continues to be true that there always
exists an element of $\Gamma$ with no invariant spinors.


\end{document}